\documentstyle[aps,epsfig,prb]{revtex}

\begin{document}

\draft

\twocolumn[\hsize\textwidth\columnwidth\hsize\csname@twocolumnfalse\endcsname

\title{Maximizing Entropy by Minimizing Area:\\ Towards a New Principle of 
Self-Organization}

\author{P. Ziherl\cite{IJS} and Randall D. Kamien}

\address{Department of Physics and Astronomy, University of Pennsylvania,
Philadelphia, PA 19104-6396}

\date{\today}

\maketitle

\begin{abstract}
We propose a heuristic explanation for the numerous non-close-packed crystal
structures observed in various colloidal systems. By developing an analogy
between soap froths and the soft coronas of fuzzy colloids, we provide a
geometrical interpretation of the free energy of soft spheres. Within this
picture, we show that the close-packing rule associated with hard-core
interaction and positional entropy of particles is frustrated by a minimum-area
principle associated with the soft tail and internal entropy of the soft coronas. We
also discuss these ideas in terms of crystal architecture and pair distribution
functions and analyze the phase diagram of a model hard-sphere--square-shoulder
system within the cellular theory. We find that the A15 lattice, known to be 
area minimizing, is favored for a reasonable range of model parameters and so it 
is among the possible equilibrium states for a variety of colloidal systems. We 
also show that in the case of short-range convex potentials the A15 and other 
non-close-packed lattices coexist over a broad ranges of densities, which could 
make their identification difficult.
\end{abstract}

\bigskip

]
\narrowtext

\section{Introduction}

Colloids are all around us. From milk to microreactors, from pie filling to
paint, and from suspensions to sieves, colloidal materials are important in a
variety of products and technologies and are the basis for a new class of
functional materials. The advent of applications based on engineered crystalline
materials, such as photonic bandgap materials,~\cite{Yablonovitch87,John87} has
put even greater focus on colloids. Both the size of colloidal particles and the
interparticle interaction are tunable, which provides the basis for the
manufacture of ordered structures with desired lattice spacings and space 
groups as well as mechanical, thermal, and electrical properties.

Over the years, considerable amounts of experimental crystallographic data have
been collected for
various colloidal systems.~\cite{Russel89} While the
tabulation of the relationship between the interparticle interaction and the
symmetry of the crystal lattice is undoubtedly valuable, it is desirable, if not
necessary, that the synthesis of colloidal crystals with specific space group be
based on analytic rather than empirical insight into the mechanisms of 
self-assembly. In principle this should be possible because the interaction
between the particles is much simpler than in atomic or molecular crystals:
colloidal particles are made of hundreds or thousands of atoms, and the
effective interaction between the particles is less specific and thus much
simpler than interatomic interactions. Thus the interaction between colloidal
particles is chiefly determined by their mechanical as opposed to chemical structure.
Here we will address those colloidal particles that are characterized by a
relatively dense core and a fluffy corona. In this case, the interparticle
potential may be approximated fairly well by the simple hard-core repulsion
dressed with a repulsive short-range interaction of finite strength.

Present theoretical understanding of the phase behavior and stability of the
various colloidal systems is quite impressive, especially in the case of charged
colloidal suspensions interacting via screened Coulomb
potentials.~\cite{Hone83,Kremer86,Rosenberg87,Rascon97} It is now well
established both by analytical approaches~\cite{Hone83,Rascon97} and by
simulations~\cite{Kremer86} that the solid part of the phase diagram of charged
colloids includes face-centered cubic (FCC) and body-centered cubic (BCC)
lattices. In the case of a convex interparticle potential -- its simplest
variant being the square shoulder interaction~\cite{Bolhuis97} -- the phase
diagram is even more complex and includes the dense and the loose FCC and BCC
lattices.~\cite{Stell72} More elaborate soft potentials such as the interaction
between star polymers lead to body-centered orthorhombic (BCO) and diamond
lattices in addition to FCC and BCC lattices.~\cite{Watzlawek99}

While extremely important, these theories do not provide a robust explanation of 
the stability of colloidal crystals, and the aim of this study is to look at the
problem from a more geometrical point of view and to capture the statistical
mechanics of colloids in a new self-organization principle. One such principle
is the maximum packing fraction rule, which states that pure excluded-volume
interactions favor an expanded close-packed structure, thereby maximizing the
configurational entropy. In the case of monodisperse hard spheres, such an
arrangement corresponds to the FCC lattice.~\cite{Mau99,Hales00} However, many
soft-sphere systems form non-close-packed lattices, including the BCC, BCO, and
diamond lattices as well as the A15 lattice~\cite{Rivier94} observed in crystals of
self-assembled micelles of some dendritic polymers.~\cite{Balagurusamy97} Can we
understand the existence of this rather loosely packed structure? Is there
another simple geometrical principle that describes the self-organization of
soft spheres and is analogous but opposing to the maximum packing fraction rule?
We pursue this question using an idealized model and find an analogy between the
soft-sphere crystals and dry soap froths, the latter being described by Kelvin's
problem of finding the minimal-area regular partition of space into cells of
equal volume. Within this framework, we propose a new principle of 
area-minimization that can favor
these loosely packed lattices of soft spheres.~\cite{Ziherl00}

Having proposed our semiquantitative and intuitive explanation for the behavior
of soft-sphere colloids, we have checked our predictions within a more rigorous
statistical-mechanical model to see whether the theoretically calculated phase
diagram includes some of these loose-packed structures -- the A15 lattice in
particular. In previous theoretical studies this structure was not considered as 
a trial state of a repulsive colloidal system, and the solid part of the phase
diagram was shared virtually exclusively by FCC and BCC 
phases.~\cite{Hone83,Kremer86,Rascon97} We scan the phase diagram for the case
of square-shoulder soft potential, the focus of other studies, and determine the 
range of widths of the soft potential where the A15 lattice is the colloidal 
ground state.

The paper is organized as follows: in Section II, we describe our model of
colloidal self-organization and establish the analogy between these systems and
soap froth. We digress to describe the Kelvin problem and its conjectured
solution. We show that this analogy leads to a second global principle of 
self-assembly -- the primary one being the principle of maximum packing entropy
-- and that the two mechanisms give rise to frustration. We apply these ideas to
a dendrimer compound (which crystallizes into the A15 lattice) as well as to
other systems. In Section III we reexamine the phase diagram of the 
hard-core--square-shoulder interaction, which captures many features of real 
colloidal systems. We analyze it within the cellular free volume approximation
using a numerical model whose main advantage is that the colloidal interaction
is treated nonperturbatively. Section IV concludes the paper.

\section{Colloidal systems as area-minimizing structures}

\begin{figure}
\centerline{\epsfig{file=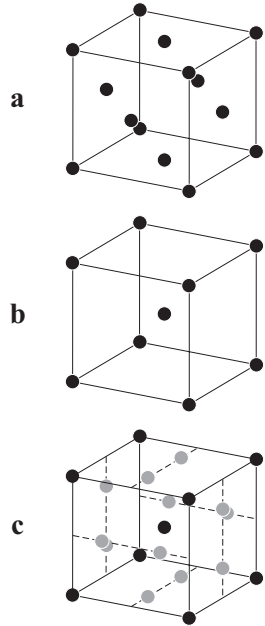}}

\vspace*{5mm}

\caption{Various lattices: (a) Face-centered cubic, (b) body-centered cubic, and (c) A15 lattices.}
\label{lattices}
\end{figure}

Recently, dendrimers composed of a poly(benzyl ether) core segments decorated with
dodecyl chains~\cite{dendrimer} were synthesized with the intention of producing
a molecule with a conical, fan-shaped architecture.~\cite{Balagurusamy97} Since
it is known that many dendrimers spontaneously self-assemble into supramolecular
clusters, the underlying rationale was motivated by the possibility of creating
spherical micellar-like objects a few nanometers in diameter. These dendrimers
do indeed form spheres which, in turn, form a crystal lattice with the 
Pm$\overline{3}$n space group, also known as the A15, Q$^{223}$, and 
$\beta$-Tungsten lattice. This lattice belongs to the cubic system, and its unit 
cell includes 8 sites which can be divided into 3 pairs of columnar sites and 2
interstitial sites. The columnar sites lie evenly spaced along the bisectors of
the faces of the unit cell and can be thought of as forming three mutually
perpendicular and interlocking columns. The interstitial sites fill out the
space between the columns: one is at the center of the cell and the other one is
at the vertex (Fig.~\ref{lattices}).

Though the structure of dendrimer crystals has been studied in great 
detail,~\cite{Percec98,Hudson99} it remains unclear why the dendrimer assemblies
should make the A15 lattice.  The micelles are nearly spherically
symmetric and very monodisperse:~\cite{Balagurusamy97} given their chemical
composition, the interaction between the micelles must be predominantly steric.
If they were of uniform density, the interaction arising from the
impenetrability of the micelles would be well-described by a spherically
symmetric hard-core potential. In this case, one might expect the spheres to
assemble into an FCC lattice to maximize their positional entropy -- because it
has the largest packing fraction, it maximizes the volume available to each 
sphere.~\cite{Mau99,Hales00} On the other hand, the A15 lattice is rather
loosely packed, with the same packing fraction as that of the simple cubic (SC)
lattice. Thus the A15 lattice is very inefficient from the point of view of the
center-of-mass entropy of each sphere.

\begin{figure}
\centerline{\epsfig{file=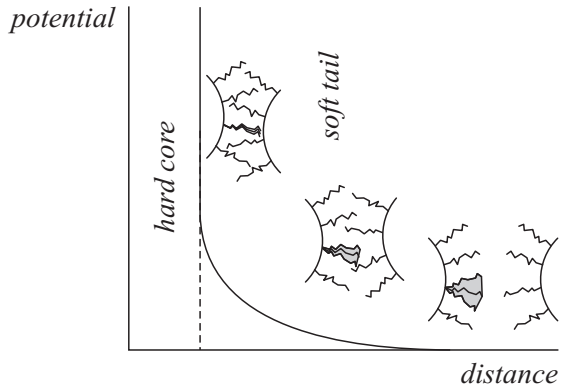}}

\vspace*{5mm}

\caption{Origin of the soft intermicellar potential: overlap between the
micelles reduces orientational entropy of chains within the corona (shaded area),
which gives rise to short-range repulsion.}
\label{inter}
\end{figure}

If the micelles were structureless hard spheres, their free energy would depend
only on their position. However, the dendrimers have a well-defined structure
consisting of a more or less compact core of benzyl ether rings and a floppy,
squishy corona of alkyl chains. In this case, the stability of a certain
arrangement of micelles does not depend only on the hard-core repulsion between
the cores but also on the interaction between the brush-like coronas: the larger
the overlap between the neighboring micelles, the more constrained the
conformations of the chains within the coronas and the smaller their
orientational entropy (Fig.~\ref{inter}). This effective interaction is, of
course, repulsive but short-ranged: at distances larger than the diameter of the
micelles, their coronas do not overlap and the interaction vanishes.

\begin{figure}
\centerline{\epsfig{file=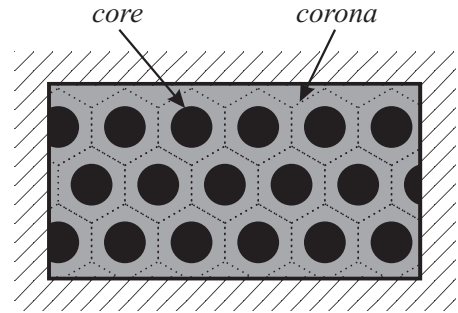}}

\vspace*{5mm}

\caption{A two-dimensional illustration of the minimal-area rule: the hard cores
are embedded in the matrix of coronas, and the volume of the latter is given by
the product of its area (dotted line) and the average separation of the cores.}
\label{minimalarea}
\end{figure}

Though the hard and the soft part of the repulsion both arise from the steric
interaction between the dendrimers, their dependence on the density is very
different. The hard cores lead to an inaccessible volume, while the matrix of
interpenetrating soft coronas leads to a softer entropic repulsion. We can
regard the matrix of coronas as a bilayer of dodecyl chains wrapped around each
hard core. The free energy of these bilayers decreases monotonically with 
thickness. However, the volume of this soft matrix is fixed by the difference 
between the total volume and that of the hard cores and can be written as 
the product of the total area ($A$) of these bilayers and their average 
thickness ($d$), so that at a given density
\begin{equation}
Ad={\rm constant.}
\end{equation}
Since the soft repulsion of the tails favors larger bilayer thicknesses $d$, the 
coronal entropy is maximized when the area $A$ is minimized, or, in other words, 
when the interfacial area between the neighboring micelles is a minimal surface 
(Fig.~\ref{minimalarea}). We should emphasize that the interface between the
micelles is a mathematical concept that embodies the membrane-like, two-dimensional 
character of the interdigitated coronas, and it does not correspond to, say, the 
position of a particular segment of the dendrimer molecule.

It is essential to note that this minimal-area principle is incompatible with 
close packing and can favor lattices other than FCC, thereby giving rise to 
frustration in these systems and a rich phase diagram. If the free energy of 
the lattice of micelles depended only on the interfacial energy, the 
system would behave as periodic membrane enclosing bubbles of equal volume -- as an ideal, dry 
soap froth. At this juncture the solution to the problem of minimizing the 
interfacial area of a set of equal volume, space-filling cells is not known. 
However, it is known that the area of the BCC ``foam'' is smaller than that of 
the FCC foam and that the A15 foam has a smaller area still.

\subsection{Kelvin's problem}

The problem of finding the ideal configuration of a soap froth was introduced by
Kelvin in 1887 while studying how light propagated in a crystal, the relation
being based on the now abandoned notion of aether. He realized that the dry soap
froth problem could be cast in the mathematically precise form: What regular
partition of space into cells of equal volume has the smallest area of cells? 
As is almost always the case, the plain and simple formulation of the 
problem is a harbinger of its complexity.

\begin{figure}
\centerline{\epsfig{file=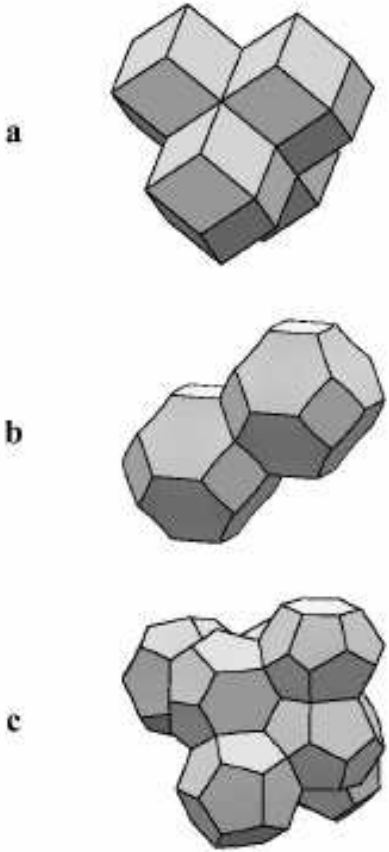}}

\vspace*{5mm}

\caption{Rhombic dodecahedron/FCC lattice (a), Kelvin's tetrakaidecahedron/BCC
lattice (b), and Weaire-Phelan minimal surface/A15 lattice (c).}
\label{bubbles}
\end{figure}

Kelvin built on Plateau's investigations of the stability of soap films,
summarized by two rules which represent mechanical equilibrium: In an
equilibrium froth, (i) adjacent faces meet at an angle of $120^\circ$ and (ii)
the edges (the so-called Plateau borders) must form a tetrahedral angle of
$109^\circ28'$. It follows that arrangements with more than 3 faces meeting at a
common edge are unstable as are junctions with more than 4 edges. At that time,
these rules did not have a theoretical background and neither did Kelvin's work.
Kelvin's approach relied on experiments based on soldered wire frames dipped
into soap solution and the observation of the evolution of the soap film spanned
by the frame.~\cite{Thomson87} Kelvin was led to the conclusion that his problem
was solved by a lattice of polygons with the topology of an orthic
tetrakaidecahedron consisting of 6 quadrilateral and 8 hexagonal faces. In more
modern terms these shapes are known as the Wigner-Seitz cell of the BCC lattice
(Fig.~\ref{bubbles}). Kelvin subsequently worked out the exact shape of the
faces and showed that to satisfy the Plateau rules, the edges of the
tetrakaidecahedra must be slightly curved and the hexagonal faces must be
somewhat nonplanar.

Though a conjecture, Kelvin's partition of space was thought to be the solution
of the problem by the mathematics community, despite the realization the
proof might be highly elusive.~\cite{Klarreich00} Indeed, these types of 
problems are notoriously difficult. For example, even the Plateau rules 
themselves were nothing but experimental facts until 1976 when they were put on
a firm theoretical footing by Taylor~\cite{Taylor76} and it was only in 1999 
that Hales proved that the regular hexagon is the solution of the 
two-dimensional variant of the Kelvin problem.~\cite{Hales01} We also note that
the related Kepler problem of packing hard spheres as densely as possible 
turned out to be equally challenging: the well-known FCC or HCP packing was 
demonstrated to be the most efficient in 1998, the proof being furnished again by 
Hales.~\cite{Hales00}

Kelvin's conjecture stood unchallenged for more than a century until 1994 when
Weaire and Phelan discovered that a froth with the symmetry of the A15 lattice
has an area smaller than BCC lattice by 0.3\%.~\cite{Weaire94a} While this may
seem to be a small difference, it is, in fact, significant: the relative
difference in the areas of Kelvin's BCC structure and the FCC-type partition of
rhombic dodecahedra (unstable as a froth because it contains vertices joining 8
edges) is 0.7\%. Thus the A15 foam is a 50\% improvement on this scale.

The A15 foam consists of 6 Goldberg tetrakaidecahedra, each with 2 hexagonal and
12 pentagonal faces, which form three sets of interlocking columns, and 2
irregular pentagonal dodecahedra at the interstices. Since it is composed of two
types of cells, the Weaire-Phelan soap froth is obviously less symmetric than
Kelvin's froth. However, it was not surprising to some that it has a rather
small surface area. The A15 lattice is derived from the polytope $\{3,3,5\}$, a
partition of positively curved space consisting of 120 regular dodecahedral
bubbles. As suggested by Kl\' eman and Sadoc, positive spatial curvature
relieves the frustration brought about by the incompatibility of the optimal
local arrangement of bubbles and the structure of flat three-dimensional space.
The polytope $\{3,3,5\}$ should be regarded as the ideal template whence the
various tetrahedrally close-packed (TCP) lattices (the layered TPC lattices are
also known as Frank-Kasper phases) are derived via decurving, {\sl i.e.}, by 
substituting some of the dodecahedral cells by bubbles with 14 ($=12$ pentagonal
$+$ 2 hexagonal), 15 ($=12$ pentagonal $+$ 3 hexagonal), or 16 ($=12$ pentagonal
$+$ 4 hexagonal) faces.~\cite{Rivier94,Kleman79,Sadoc99} Cells with 13 faces are
forbidden for topological reasons, and those with more than 16 faces are
dynamically unstable.~\cite{Rivier94}

There are 24 known ways of decurving the polytope $\{3,3,5\}$ and thus 24 TCP
crystal lattices.~\cite{Rivier94} With 2 types of bubbles and 8 bubbles per unit
cell, the A15 lattice is among the simplest: the unit cell of the most
complicated structure, the so-called I lattice, consists of 228 bubbles that
include all 4 types of bubbles.~\cite{Rivier94} Given the rationale for the
success of the A15 lattice, it is not unreasonable to expect that other TPC
lattices may have an even smaller surface area. Indeed, the discovery of Weaire
and Phelan renewed the interest in the field. Subsequently some of the remaining
TCP-type soap froths have been studied,~\cite{Weaire97} facilitated by Surface
Evolver, a remarkable software package developed by Brakke.~\cite{Brakke92} In
addition, other classes of periodic and quasiperiodic partitions based on
Kelvin's BCC and Williams' body-centered-tetragonal (BCT) 
bubbles have been suggested.~\cite{Glazier94} At
this time, however, the A15 foam has not been bested -- it stands as the
tentative solution of Kelvin's problem.

The connection between the observed crystal structure and the soft part of the
intermicellar potential should now be clear: if it were absent, the hard cores
would favor an FCC lattice, but instead the equilibrium structure is a different
lattice with a larger bulk free energy but a smaller surface free energy. The
only variables in this model are the parameters of the surface interaction,
which we will estimate roughly. In view of the experimental data on soft
spheres, we will compare the FCC, BCC, and A15 lattices and show that in the
case of the dendrimer micelles, it is reasonable to expect that the 
area-minimizing A15 lattice is favored over the close-packed FCC lattice.

\subsection{Bulk free energy}

Within the framework of the two competing ordering principles, we propose a
simple and approximate theory of colloidal crystals where the bulk and the
surface terms are coupled only through the constraint of fixed volume. As far as
the bulk free energy is concerned, the micelles will be treated as hard spheres
of diameter $\sigma$, and the surface interaction will be calculated as if the
matrix of coronas were a thin structureless layer so that we may neglect the
effects of curvature.

We start our analysis with the bulk term. The configuration integral of a 
one-component classical system of $N$ particles confined to a volume $V$ is
\begin{equation}
Z=\frac{1}{\lambda^{3N}N!}\int_V\prod_{i=1}^N{\rm d\bf r}_i
\,\exp\biglb(-U({\bf r}_1,{\bf r}_2,\ldots{\bf r}_N)/k_{\scriptscriptstyle
B}T\bigrb),
\end{equation}
where $\lambda=\sqrt{h^2/2\pi mk_{\scriptscriptstyle B}T}$ is the thermal de
Broglie wavelength and $U({\bf r}_1,{\bf r}_2,\ldots{\bf r}_N)=\frac{1}{2}
\sum_{i,j=1}^Nu({\bf r}_i,{\bf r}_j)$ is the total interaction energy consisting 
of pairwise interactions $u({\bf r}_i,{\bf r}_j)=u(|{\bf r}_i-{\bf r}_j|)$. 
Despite the simplicity of the hard-core potential, $Z$ cannot be calculated
analytically, and one must resort either to numerical approaches, such as Monte
Carlo simulations, or to approximate analytical methods. The latter are more
appropriate for our purposes since we are seeking a simple, heuristic
explanation of crystal structure. When considering crystal phases we can assume
that the particles are localized within cells formed by their neighbors and thus
the configuration integral breaks up into $N$ single-particle integrals so that
\begin{equation}
Z\approx\left(\lambda^{-3}\int_{V_0}{\rm d}{\bf r}\,\exp(-u_{\rm eff}({\bf r})/
k_{\scriptscriptstyle B}T)\right)^N,
\label{cellmodel}
\end{equation}
where $V_0$ is the volume of the cell (usually the Wigner-Seitz cell), 
$u_{\rm eff}(\bf r)$ is the effective potential felt by the particle, and $N!$ has
been absorbed by factorization of the partition function. This approximation is 
valid only for lattices where all sites are equivalent; if the unit cell of the 
crystal consists of inequivalent sites, Eq.~(\ref{cellmodel}) can be embellished 
accordingly.

As natural as this approximation may seem for a periodic arrangement of
particles in a crystal, it neglects any correlated motion of neighbors and the
associated communal entropy.~\cite{Hill56,Barker63} Nevertheless, the cellular
model often gives quantitatively accurate results that are in good agreement with
more complete numerical simulations. For hard spheres, where the cellular theory
becomes exact in the high-density limit, the agreement with Monte Carlo results
is in fact excellent.~\cite{Barker63,Curtin87} In this case, each particle is
assumed to be uniformly smeared over its reduced Wigner-Seitz so that it cannot
overlap with its neighbors. As a result, the effective potential is quite
simple: 0 within the volume the center of mass is allowed to trace out (the 
so-called free volume) and infinite otherwise. Thus
\begin{equation}
Z\approx\left(\lambda^{-3}\int_{V_{F}}{\rm d}{\bf r}\right)^N=\left(V_{F}/
\lambda^3\right)^N,
\label{cellmodel3}
\end{equation}
where $V_{F}$ is the free volume whose shape reflects the symmetry of the
lattice.\cite{Barker63} We note that temperature drops out of this problem and
that the interaction is entirely entropic.

Within the cellular free-volume theory, the free energy of hard spheres becomes
a purely geometrical issue. All one has to do is to calculate the volume of the
Wigner-Seitz cell after a layer of thickness $\sigma/2$ has been peeled off of
its faces, which is the volume accessible to the particle's center of mass. In
the FCC lattice, the free volume has the shape of a rhombic dodecahedron just as
the Wigner-Seitz cell.~\cite{Kittel53} On the other hand, in the BCC lattice it
remains an orthic tetrakaidecahedron only at rather low densities far below the
freezing point. At higher densities, where the hard spheres form a solid phase,
the square faces become absent rendering the free volume a regular octahedron.
For the FCC and BCC lattices, the bulk free energy per dendrimer micelle reads
\begin{equation}
F^X_{\rm bulk}=-k_{\scriptscriptstyle B}T\ln\Bigglb(\alpha^X
\left(\frac{\beta^X}{n^{1/3}}-1\right)^3\Biggrb),
\label{fccbcc}
\end{equation}
where $X$ stands for FCC or BCC, $n=\rho\sigma^3$ is the reduced number density
($\sigma$ being the hard-core diameter), and the term
$-3k_{\scriptscriptstyle B}T\ln(\sigma/\lambda)$ which is independent of the
lattice structure has been dropped. The coefficients $\alpha^{\rm FCC}=2^{5/2}$
and $\alpha^{\rm BCC}=2^23^{1/2}$ reflect the shape of the free volume, and
$\beta^{\rm FCC}=2^{1/6}$ and $\beta^{\rm BCC}=2^{-2/3}3^{1/2}$ specify their
size.

\begin{figure}
\centerline{\epsfig{file=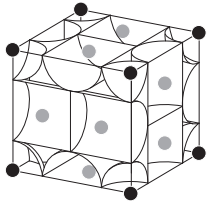}}

\vspace*{5mm}

\caption{Approximate analytical model of the bulk free energy for the A15
lattice: free volumes of columnar and interstitial sites are replaced by
cylinders and spheres, respectively.}
\label{spheres}
\end{figure}

\begin{figure}
\centerline{\epsfig{file=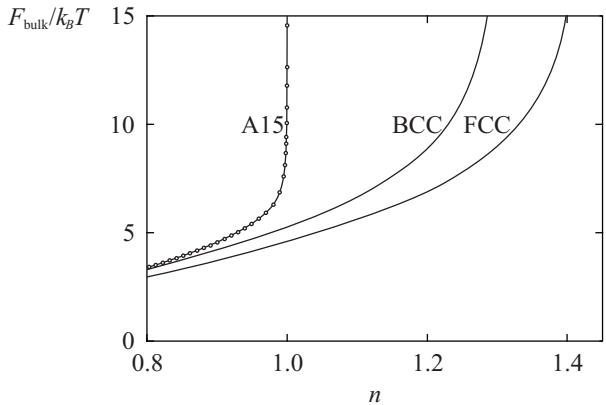,width=80mm}}

\vspace*{5mm}

\caption{Bulk free energies of hard-core particles arranged in FCC, BCC, and A15
lattice as calculated with the free-volume theory. Solid lines correspond to 
Eqs. (\ref{fccbcc}) and (\ref{a15}), and circles are numerical results. }
\label{fbulk}
\end{figure}

The free volume of the A15 lattice, which includes two types of sites, is a bit
more complicated. As determined by the Voronoi construction subject to the
constraint that all cells have equal volume, the shapes of the free volumes are
irregular pentagonal dodecahedra and tetrakaidecahedra with two hexagonal and
twelve pentagonal faces rather than regular polyhedra. Their volumes cannot be
expressed in an amenable analytical form, and thus we have calculated them
numerically for a range of densities. However, we note that the exact free
volumes can be approximated very well by replacing the dodecahedra and
tetrakaidecahedra by spheres and cylinders, respectively (Fig.~\ref{spheres}).
This {\sl ansatz} takes into account that the free volume of a columnar site
changes anisotropically with density and approaches a flattened shape in the
close-packing limit, whereas the shape of the free volume of interstitial sites
does not depend on density. We also introduce two adjustable parameters that
quantify the fact that the actual volumes of the Wigner-Seitz cells are larger
than the volumes of spheres and cylinders, which leave empty voids between them.
Given that the ratio of columnar and interstitial sites is 3:1, this leads to
the average bulk free energy per micelle of
\begin{eqnarray}
F_{\rm bulk}^{\rm A15}&=&-k_{\scriptscriptstyle B}T\left[\frac{1}{4}\ln
\Bigglb(\frac{4\pi S}{3}\left(\frac{\sqrt{5}}{2n^{1/3}}-1\right)^3\Biggrb)
\right.\\ \nonumber
&&+\left.\frac{3}{4}\ln\Bigglb(2\pi C\left(\frac{\sqrt{5}}{2n^{1/3}}-1
\right)^2\left(\frac{1}{n^{1/3}}-1\right)\Biggrb)\right].
\label{a15}
\end{eqnarray}
This formula best agrees with the numerical results for $S=1.638$ and $C=1.381$,
where the relative deviation from the true bulk free energy is below 0.1\% at
densities higher than $n\approx0.8$ and not significantly larger at lower
densities. $F_{\rm bulk}^{\rm FCC}$, $F_{\rm bulk}^{\rm BCC}$, and
$F_{\rm bulk}^{\rm A15}$ are plotted in Fig.~\ref{fbulk}.~\cite{Ziherl00}

\begin{figure}

\centerline{\epsfig{file=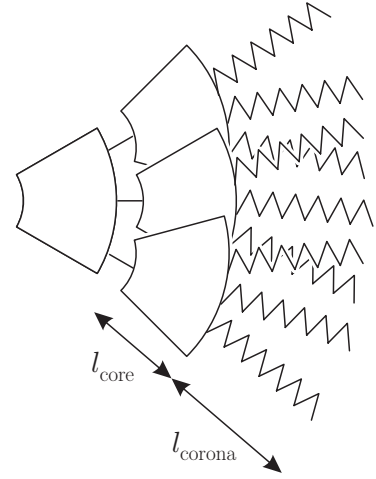}}

\vspace*{5mm}

\caption{Structure of dendrimers that form the A15 lattice: 1st generation
generation dendrimer consists of a core segment and three dodecyl chains, 2$^{\rm nd}$ 
generation dendrimer includes three 1$^{\rm st}$ generation dendrimers attached to a core 
segment, and so on. The length of the core segment and the dodecyl tail is 
$l_{\rm core}=0.6$~nm and $l_{\rm corona}=1.4$~nm, respectively. Shown here is a 
2$^{\rm nd}$ generation dendrimer.}
\label{corechain}
\end{figure}

\subsection{Surface free energy}

Having calculated the bulk free energy for the three lattices, we now turn to 
the surface free energy, which requires a specific model of the soft interaction
between the particles. To determine which model is the most appropriate, we should
first find out whether the overlap of the neighboring particles is large or small.
At this point, the analysis becomes somewhat less general, because the overlap
differs from one system to another. Being interested primarily by the stability of 
the loosely-packed crystal lattices, we now focus on the dendrimers that form
the A15 crystal.~\cite{Balagurusamy97} The relevant quantitative data include 
measurements of the radii of dendrimer molecules and the lattice constants of the 
micellar crystals. Since the dendrimers consist of 2, 3, or 4 generations of
branching benzyloxy segments crowned by dodecyl chains with bare radii of
$\sigma_{\rm bare}/2=2.6,$ 3.2, and 3.8~nm, respectively, we can deduce that the
lengths of the benzyloxy core segment and the dodecyl chain are $l_{\rm core}=
0.6$~nm and $l_{\rm corona}=1.4$~nm (Fig.~\ref{corechain}). The effective diameter 
of the micelles $\sigma_{\rm eff}$ can be calculated from the lattice constant $a$. 
According to Fig.~\ref{lattices}, $\sigma_{\rm eff}=a/2$. This gives $\sigma_{\rm 
eff}=3.4,$ 4.0, and 4.2~nm. Obviously, $\sigma_{\rm eff}$ is considerably smaller 
than $\sigma_{\rm bare}$ for all generations. Actually, the most conclusive 
information can be extracted from 4$^{\rm th}$ generation data: 2$^{\rm nd}$ and 3$^{\rm rd}$ 
generation 
micelles probably have an empty center and their true diameter is most likely 
larger than $\sigma_{\rm bare}$. These data show that the effective diameter of the 
hemispheric 4$^{\rm th}$ generation micelles, 4.2~nm, does not exceed the diameter of the 
their benzyloxy core, $8l_{\rm core}=4.8$~nm. Thus we can conclude that the 
hard-core diameter of the monodendrons must be smaller than the diameter of the 
benzyloxy core. Thus not only is there a considerable overlap between the coronas, 
but also the dodecyl chains penetrate into the core itself. This indicates that the 
interfacial effects, related to the limited orientational entropy of the chains, 
are important in the dendrimer system.

In the absence of a quantitative insight into the intermicellar potential -- 
such as a direct measurement of the interaction via optical 
trapping~\cite{Crocker94} -- we model the interaction of the interpenetrating 
dodecyl chains as the interaction between grafted polymer brushes.~\cite{Milner88} 
In the high-interdigitation limit, which is certainly applicable in our case, the 
free energy of a compressed brush reduces to the excluded-volume repulsion of the
chains. An argument in the spirit of Flory theory gives 
\begin{equation} 
F_{\rm surf}=\frac{\ell N_0k_{\scriptscriptstyle B}T}{h}=\frac{2\ell 
N_0k_{\scriptscriptstyle B}T}{d}, 
\end{equation} 
where $\ell$ is a parameter with the dimension of length, which determines the
strength of repulsion, $N_0$ is the number of chains per micelle, and $h$, the
thickness of the single corona, is half the average thickness of the
interdigitated matrix of the chains $d$. This approximation neglects some
details of the actual interaction, most notably the curvature of the dodecyl
brushes in the coronas. However, these effects are subdominant when the
dendrimers are packed very closely. In fact, it is known that the density is
nearly constant in the volume occupied by the chains,~\cite{Balagurusamy97}
and we thus expect that our model should provide a robust description of the
system.

The thickness $d$ of the coronal matrix depends on the density since the dodecyl
bilayer must fill the space between the hard cores.~\cite{Ziherl00} Thus
\begin{equation}
d=\frac{2(n^{-1}-\pi/6)\sigma^3}{A_M},
\end{equation}
where $A_M$ is the interfacial area per micelle. Since the area is proportional
to the square of the lattice constant, the surface free energy per micelle reads
\begin{equation}
F^X_{\rm surf}=\frac{\ell N_0k_{\scriptscriptstyle
B}T}{\sigma}\frac{\gamma^Xn^{-2/3}}{n^{-1}-\pi/6},
\label{surf}
\end{equation}
where $\gamma^X$ is the coefficient determined by the symmetry of the lattice
and defined by $A_M=\gamma^X\sigma^2n^{-2/3}$. Lower values of $\gamma^X$
correspond to more area-efficient partitions. The typical magnitude of this
coefficient is set by the simple cubic (SC) lattice, which gives
$\gamma^{\rm SC}=6$. The more efficient area-minimizing lattices have smaller values of
$\gamma^X$: $\gamma^{\rm FCC}=2^{5/6}3=5.345$, $\gamma^{\rm BCC}=5.306$, and
$\gamma^{\rm A15}=5.288$.~\cite{Weaire94a} $\gamma^{\rm BCC}$ and
$\gamma^{\rm A15}$ can only be computed numerically, e.g., using Surface
Evolver.~\cite{Brakke92} As a comparison, the ultimate lower bound for $\gamma$
corresponds to a sphere (which, of course, is not a space-filling body):
$\gamma^{\rm sphere}=2^{2/3}3^{2/3}\pi^{1/3}=4.836.$

Combining Eqs.~(\ref{fccbcc}), (\ref{a15}), and (\ref{surf}), we arrive at a
single-parameter ($\ell$) free energy of micellar crystals:
\begin{equation}
F^X=F_{\rm bulk}^X+F_{\rm surf}^X.
\end{equation}
Before we calculate the minimal strength of the soft repulsion necessary to
stabilize the area-minimizing structures, we need to determine the density of
the crystal. Not knowing the hard-core radius of the micelles, we can only
provide an order-of-magnitude estimate of $\ell$, and for this purpose, it is
sufficient to note that unless the observed A15 lattice approaches the packing
limit of $n=1.0$, its bulk free energy per micelle is roughly
$1k_{\scriptscriptstyle B}T$ larger than $F_{\rm bulk}^{\rm BCC}$ and
$2k_{\scriptscriptstyle B}T$ larger than $F_{\rm bulk}^{\rm FCC}$. To be
concrete, we compare the total free energies at $n=0.95$, which is well within
the high-density regime but not quite at the close-packing limit and thus
consistent with the structural data. At this density, the BCC lattice becomes
more favorable than FCC at $\ell\gtrsim0.05 \sigma$ whereas the BCC-A15
transition occurs at $\ell\approx0.15\sigma$. Given that there are $N_0=162$
dodecyl chains per 3rd and 4th generation micelle, these values of $\ell$
correspond to an entropy of about $0.5k_{\scriptscriptstyle B}$ and 
$1.5k_{\scriptscriptstyle B}$ per chain, respectively. In other words, this 
means that if the overlap of the micelles is so large that the decrease of 
orientational entropy of a chain due to interdigitation reaches 
$0.5k_{\scriptscriptstyle B}$ and $1.5k_{\scriptscriptstyle B}$, the differences 
between the surface entropies of the BCC and FCC lattices and the A15 and BCC 
lattices overcome the corresponding differences of the bulk entropies and thus
favor the area-minimizing structures.

These values are physically reasonable and of the correct order of magnitude.
When unrestricted by other chains, each chain has a few orientational and
conformational degrees of freedom. Thus its entropy is a few 
$k_{\scriptscriptstyle B}$, and so it can easily loose 
$0.5k_{\scriptscriptstyle B}$, $1.5k_{\scriptscriptstyle B}$, or more entropy 
upon interdigitation. If we plug these numbers back into the free energy, we 
learn one more thing about these systems: their energetics is controlled
primarily by the surface term, which is a direct consequence of the three
center-of-mass degrees of freedom associated with the hard cores being greatly
outnumbered by the several hundreds of internal degrees of freedom associated
with the dodecyl chains. In other words, in the dendrimer system the 
``squishiness'' of the particles wins over their hard cores.

\subsection{A new paradigm}

Our proposed model provides a novel way of looking at the self-organization of
colloidal crystals. By complementing the well-known close-packing rule, which
controls the stability of hard spheres, with the minimal-area rule, which stems
from the additional short-range repulsion between the particles, we have shown
that the equilibrium ordered structures are a result of frustration between two
incompatible requirements. This picture implies that if the corona is thin
compared to the core, the colloids will behave as hard spheres and form a 
close-packed lattice such as FCC, but as it grows thicker, an area-minimizing
structure -- the A15 lattice -- should be observed. In between the two extremes,
there may be a spectrum of lattices that neither maximize the packing fraction
nor minimize the interfacial area but represent a reasonable compromise for a
given intermicellar potential. However, in some systems coexistence of the FCC
and A15 lattices could be energetically preferable to an intermediate structure
such as BCC, provided that the density is not too high to destabilize the A15
lattice.

This observation is consistent with the experimentally determined structures
found in other colloidal materials, such as crystals of aurothiol particles
consisting of a gold crystallite core and covered by about 50 $n$-alkylthiols
where $n=4$, $6$, or $12$. This system is remarkably close to the dendrimer
system in both size and structure. The diameter of the metallic cores can be
varied from 1.6 to 3.1~nm, and the length of the (fully extended) coronal chains
is between 0.6 to 1.56~nm. The alkylthiol chains were adsorbed to the gold core
with the sulfur atom, leaving the outer part of the corona chemically identical
to the dendrimer coronas. Depending on the relative size of the corona with
respect to the diameter of the core, either the FCC or BCC lattice was observed.
In addition, in some samples the BCT lattice was found as a moderately
anisotropic variation of the BCC lattice with $c/a\sim1.15$. (We will see later
on that this lattice is not really unexpected although it departs somewhat from
our model: in a BCT crystal, both the bulk and surface entropy are larger than
their BCC counterparts.)

A particularly interesting feature of this system is that the size of the core
can be varied continuously, so that the FCC-BCC transition can be located very
precisely. For the range of core radii explored in the study,~\cite{Whetten99}
an FCC-BCC transition was found in particles covered with hexylthiol chains: the 
FCC lattice is stable if the ratio of the thickness of corona and the core 
radius is smaller than about 0.73, and BCC or BCT lattices are observed 
otherwise. The structural parameters of the aurothiol particles were measured in 
some detail and if we identify the hard core with the gold nanocrystal, we find
that the reduced density at the transition is about $n_{\rm FCC-BCC}=0.4$. 
Proceeding along the same lines as before, this gives an entropy decrease of 
about $0.05k_{\scriptscriptstyle B}$ for each of the approximately 150 chains in 
the corona. Although still reasonable, this estimate is by an order of 
magnitude smaller than in the dendrimer case. However, we note that because of 
the high coverage of the gold cores with the adsorbed alkylthiols, the effective 
hard-core diameter of the particles is most likely larger than the gold core 
diameter. In this case, the reduced density at the transition would be larger, 
and this would lead to a larger entropy decrease per alkylthiol chain. 
Nonetheless we can still combine this figure with its dendrimer counterpart to 
bound the value of the entropy per coronal chain at the FCC-BCC transition between 
0.05 and $0.5k_{\scriptscriptstyle B}$, which can serve as an estimate for future 
studies.

Similar behavior was discovered in polystyrene-polyisoprene diblock copolymers
dispersed in decane, a solvent preferential for the 
polyisoprene.~\cite{McConnell93,McConnell96} These diblocks spontaneously form
micelles with a polystyrene core and polyisoprene corona. Indeed, micelles based 
on diblocks with a core segment containing from 1.5 to 2 times as many monomers
as the coronal segment crystallize into an FCC lattice.~\cite{McConnell96} On 
the other hand, in copolymers made of blocks with the same number of monomers 
the BCC lattice was observed. Although the A15 lattice was not seen in this 
system, we conjecture that it could be found in copolymers with the polyisoprene 
block sufficiently longer than the polystyrene block.

While rare in crystalline systems, we mention that the A15 structure is not
uncommon in lyotropic systems, typically containing lipids dispersed in water
matrix. These systems self-arrange in either direct or inverted micelles, which
are known to form a variety of cubic 
structures.~\cite{Charvolin88,Luzzati96,Mariani94,Clerc96} However, the
existence of the A15 lattice in lyotropic systems should not be as surprising as
in colloidal systems. The cohesive force of these micelles is the hydrophobic
interaction which means that their size is not as well-defined as in colloids.
If there can be several types of micelles the A15 lattice is not hard
to assemble. The other difference with respect to colloids is that the water
matrix that encloses the micelles is truly fluid and thus much more similar to
actual wet soap froths with relatively large liquid content. At the same time,
in the lyotropic systems the effective intermicellar potential may not be
necessarily dominated by steric effects, implying that the stability of a
particular ordered micellar structure can depend on the chemical composition of 
lipids. 

Specific effects like these can not easily be incorporated into our framework:
our goal was not to develop an elaborate description that would cover the many
details that are captured in more or less involved theories, such as 
self-consistent field theory~\cite{Matsen97} and various Monte Carlo 
schemes.~\cite{Watzlawek99,Dotera99} Instead, we have proposed a heuristic model
that describes the essential physics of many colloids and clearly exposes the
frustration that is introduced by dressing hard spheres with a soft repulsive
interaction. This model provides an explanation of the results of the more
rigorous theories with the added advantage that it is very simple, yet can give
reasonable semiquantitative predictions about the stability of the different
lattices. Obviously, it can be improved by introducing a more refined model of
the interfacial interaction which would account for the curvature of the coronal
matrix, the strain of the coronas into the interstitial regions, solvent
effects, and related phenomena.

As useful as it might be, like all coarse-grained approaches, the proposed
theory~\cite{Ziherl00} has some limitations. Our model is tailored for a broad
yet specific class of colloidal particles that interact by short-range 
potentials, and it may not be suitable for systems characterized by long-range
forces, such as unscreened or partly screened electrostatic forces. The range of
potential is determined by the requirement that the nearest-neighbor interaction
be much stronger than the interaction with the rest of the particles; otherwise
we could not define the concept of the interface. Additionally, we have
decoupled the bulk and interfacial free energies, or the hard and soft parts of
the potential. We suspect that a limit of the full problem exists in which the
minimum-area principle is a mean-field approximation. 

In order to gain insight into colloidal crystal structure and to strengthen the
case for our new principle, we will, in the next Section, calculate the phase
diagram for a certain interparticle potential, and identify the range of
stability of the A15 lattice using a ``single-particle'' Monte Carlo algorithm.
The significance of these results is twofold: at the qualitative level, they
suggest that the phase diagram of soft spheres can include a variety of solid
phases, not just FCC and BCC lattices, and at the quantitative level they will
provide guidance for a more complete numerical analysis of the system.

\section{Short-range potentials and solid-solid phase transitions}

\subsection{Crystal architecture}

\begin{figure}

\centerline{\epsfig{file=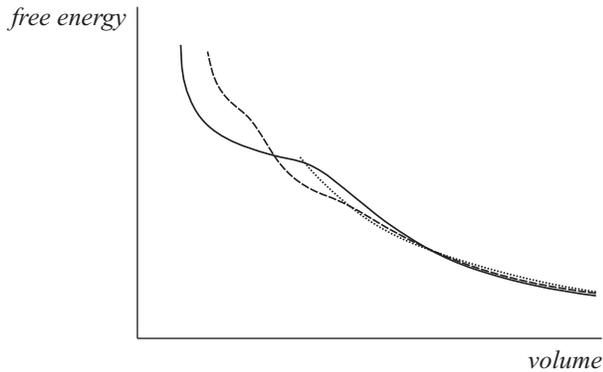,width=80mm}}

\vspace*{5mm}

\caption{Schematic of the free energy profile for a model hard-core--soft-shoulder 
potential. The short range of the shoulder makes the free energy a nonconvex 
function of volume, and thus induces either isostructural or structural transitions 
in the solid phase. Solid line: FCC; dashed line: BCC; dotted line: A15.}
\label{feschematic}
\end{figure}

The FCC lattice, which, as we know now for certain,~\cite{Hales00} is the
closest regular packing of hard spheres with cubic symmetry, with each site
enclosed completely by the 12 equidistant neighbors. In the case of a purely
hard core interaction, the free energy consists solely of the entropic
contribution and is a convex function of volume. To model the effect of the soft
coronas in the systems we are discussing, we add to the hard core a soft
potential of range $\delta$. Note that we are replacing an entropic interaction
with an energetic term which should make no difference to our argument but is
more convenient in our calculational scheme. At densities low enough so that the
average distance between the particles is much larger than $\delta$, the overlap
of the coronas is small and so the free energy is dominated by the entropy, as
if the extra potential were not there. On the other hand, in the high-density
limit where the separation between the hard cores is smaller than $\delta$, each
particle feels the soft potential of all neighbors. If the potential is flat
enough, the average field may in fact not change dramatically with position and
then the particle will behave essentially as if it were a hard sphere moving in
a constant potential, its free energy being given by the entropic contribution
of the hard-core interaction shifted by the energy of the overlapping coronas.
For certain forms of the soft potential, the transition from the low-density to
high-density regime can be rather abrupt (Fig.~\ref{feschematic}). Upon
compression, the free energy of the colloid will then change quite sharply and
may no longer be a convex function of volume. In this case, the low-density and
high-density behavior will be separated by a region of coexistence between
expanded and condensed FCC lattices, implying an isostructural transition
between them.~\cite{Bolhuis97}

Coexistence between the expanded and condensed FCC phases is not the only
possibility: other crystal lattices may exist in this density regime where it
may be favorable for the spheres to be configured in a structure with a somewhat
higher entropy but lower energy. For example, the BCC lattice has a more open
architecture compared to FCC where the ``cage'' of each particle is defined by
the 12 nearest neighbors. In the BCC lattice, the 8 nearest neighbors do not
enclose the particle completely: as the Wigner-Seitz cell shows us,
even in the hard-core limit, each particle 
also
interacts with the 6 next-to-nearest neighbors. At any fixed density, the 8 
neighbors from the first coordination shell are closer to the particle than in the 
FCC crystal, but the 6 second-shell neighbors are further away. As a result, BCC 
hard spheres have a lower entropy than the FCC hard spheres, but the additional 
soft potential can stabilize the BCC lattice. The more structured arrangement of 
sites implies that upon compression, the energy of the BCC crystal increases more 
gradually than its FCC counterpart: the central particle first overlaps with the 
first-shell neighbors alone and in this density regime the BCC free energy can be 
lower than the free energy of the coexisting expanded and condensed FCC phases, 
provided that the soft potential overlaps with all 12 neighbors of the FCC lattice. 
At somewhat higher densities the particles in the BCC lattice interact with all of 
the 14 neighbors and then both the entropic and energetic terms disfavor the BCC 
lattice. Similarly, the stability of the A15 lattice and other more structured 
phases is facilitated by their even smoother, less step-like free energy as a 
function of volume.

\begin{figure}
\centerline{\epsfig{file=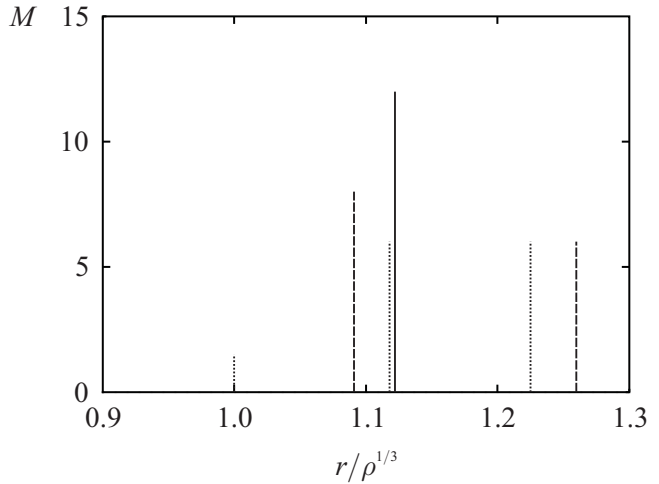}}

\vspace*{5mm}

\caption{Structure of the FCC (solid line), BCC (dashed line), and A15 lattice
(dotted line) represented by the radii of a few nearest coordination shells
weighted by the corresponding coordination number $M$. (If peaks were multiplied
by $\delta$-functions, this would be the $T=0$ pair distribution function
integrated over the solid angle.) FCC is the most closed lattice, each particle
having 12 equivalent neighbors, whereas BCC and A15 are more open structures
with 2 and 3 inequivalent neighbor sites, respectively.}
\label{rdf}
\end{figure}

This view of the energetics is closely related to the structure of the pair 
distribution function of a particular lattice. As long as the interaction between
the particles is short-range, one only has to consider the pair distribution
function at short separations, which directly reflects the shape of the 
corresponding Wigner-Seitz cell. If the distribution function has a single peak as 
in the highly symmetric FCC crystal, a soft short-range repulsion between the
particles can destabilize the lattice with respect to other lattices with more
complex pair distribution functions. As noted above, these include the BCC lattice
with two peaks at short distances (Fig.~\ref{rdf}) but also the anisotropic variants
such as the BCT lattice~\cite{Whetten99} where the 6 BCC second-shell sites are
split into a subshell of 4 sites and a subshell of 2 sites, and the BCO lattice~\cite{Watzlawek99} where the fourfold BCT 
next-to-nearest-neighbor subshell is further split in two subshells. (Curiously, 
the anisotropic version of the FCC lattice -- the face-centered orthorombic (FCO) 
structure has not yet been observed in a colloidal or related system.)

This argument can be extended to include other, less symmetric lattices. An
example is the A15 lattice. By considering the pair distribution function, we
see that, on average, each site interacts with 13.5 neighbors: through the
faces of their Wigner-Seitz cells, the columnar and interstitial
sites interact with 14 and 12 neighbors, respectively, and there are 3 times as many
columnar sites as the interstitial sites. Of these 13.5 neighbors, 1.5 particles are
closest to the central particle, followed by two shells of 6 particles at larger
distances~(Fig.~\ref{rdf}). While it is obvious that this lattice must have a
smaller entropy than BCC, for a suitably chosen soft potential it could have a
lower net free energy. An extreme case of the tradeoff between entropy and
energy is the diamond phase observed in a numerical study of packing of star
polymers which interact with an extremely soft potential.~\cite{Watzlawek99} The
diamond lattice with 4 nearest and 12 next-to-nearest neighbors is clearly
unfavorable from the entropic point of view, but apparently the pair
distribution function is so strongly peaked at next-to-nearest neighbors that
for certain types of short-range interparticle potentials the energy of the
diamond lattice is very low.

While these ideas provide additional perspective and complement our foam model
described above, it is not completely clear whether the structural
characteristics captured by the pair distribution function can be easily related
to the minimal-area principle and Kelvin problem. We will not pursue this
interesting question here; instead, we now turn to statistical mechanics of soft
spheres.

\subsection{Phase diagrams}

The phase behavior of soft-sphere systems has had ongoing study over the years.
A variety of purely repulsive forces have been considered, including power law
potentials as the high-temperature limit of the van der Waals 
interaction,~\cite{Hoover70} screened Coulomb potentials with~\cite{Meijer97} or
without a hard core,~\cite{Kremer86,Hone83} and square shoulder 
potentials~\cite{Bolhuis97,Rascon97,Lang99} as well as their 
variants.~\cite{Velasco00} With its flat plateau and as sharp a cutoff as
possible, the square-shoulder potential is very sensitive to the structure of
the pair distribution function and thus the quintessential short-ranged
interaction. This is the reason why it is used so often to analyze solid-solid
transitions.  We shall use the square-shoulder interaction in the
following, in order to provide a comparison with earlier studies.

\begin{figure}

\centerline{\epsfig{file=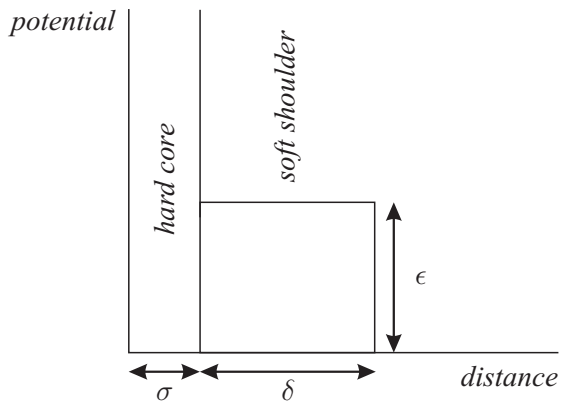}}

\vspace*{5mm}

\caption{Hard-core--soft-shoulder potential}
\label{hcss}
\end{figure}

The hard-core square-shoulder (HCSS) system is characterized by a pairwise
potential of the form
\begin{equation}
u(r) = \left\{
\begin{array}{lcl}
\infty & &r<\sigma\\
\epsilon & & \sigma\leq r<\sigma+\delta\\
0 &&r>\sigma+\delta\\
\end{array}
\right.
\end{equation}
where $\delta$ and $\epsilon>0$ are the thickness and the height of the
shoulder, respectively (Fig.~\ref{hcss}). This system has already been 
studied~\cite{Rascon97,Velasco98} by using density-functional perturbation
theory, which treats the behavior of the system primarily by the hard-core
interaction and treats the additional soft potential as a perturbation. If the
free energy and the structure of the reference state ($F_{\rm ref}$) are known,
the total free energy may be expanded to linear order in the perturbative 
term~\cite{Hansen86}:
\begin{equation}
F[\rho({\bf r})]=F_{\rm ref}[\rho({\bf r})]+\frac{1}{2}\int{\rm d}{\bf r}_1\,
{\rm d}{\bf r}_2\,\rho_{\rm ref}^{(2)}({\bf r}_1,{\bf r}_2)\,\phi(r_{12}),
\end{equation}
where ${\bf r}={\bf r}_2-{\bf r}_1$ describes the relative position of the two
particles, $\rho_{\rm ref}^{(2)}({\bf r}_1,{\bf r}_2)$ is the pair distribution
function of the reference state, and $\phi_{\rm pert}$ is the perturbative part
of the potential. The inputs of this scheme are the pair distribution function
and the free energy of the reference system. Typically, the free energy of the
solid HC phases are described within the so-called cellular free volume 
theory~\cite{Barker63,Hill56} and the corresponding pair distribution is 
calculated from the Gaussian one-particle density, whereas the fluid free 
energy and pair distribution are described by the Carnahan-Starling formula and
the Verlet-Weis formula, respectively.~\cite{Hansen86}

Within perturbation theory, it has been found that the solid part of the phase
diagram can indeed be very complex, and that its topology depends very
delicately on\break 

\noindent
the parameters of the potential. For relatively narrow shoulders,
the phase diagram differs from its hard-core counterpart only in the expanded
FCC--condensed FCC transition that occurs at densities near the close-packing
density and terminates at a critical point. As the shoulder becomes broader, the
region of coexistence between the expanded and condensed phases shifts towards
lower densities. Eventually, the expanded FCC structure is replaced in part by
the BCC at intermediate temperatures, {\sl i.e.}, at temperatures high enough so
that the shoulder does not appear very high, yet low enough so that it is not
irrelevant. According to this analysis, the transition between the fluid and
condensed FCC phase can be either direct or indirect, the intervening phases
being BCC, expanded FCC, or both. Similar results were obtained for a sloped
shoulder and a square shoulder followed by a linear ramp, except that these
potentials give rise to several new features of the phase diagram, such as the
expanded BCC--condensed BCC transition and a triple point. These predictions are
consistent with the limited Monte Carlo results available,~\cite{Bolhuis97} but
most of the features remain to be verified by rigorous numerical studies.

The perturbation analysis has provided valuable information on the behavior of
HCSS spheres, but its predictive power is limited to high temperatures where the
expansion is valid. Although some additional information can be obtained by
interpolating between the high-temperature and zero-temperature data, this
leaves out the most interesting part of the phase diagram. A natural extension
of this approach is to calculate the configuration integral nonperturbatively.
This can be done, for example, within a mean-field approximation in the cellular
model where each particle is assumed to move independently in the average field
of its neighbors. Such an approximation can be used within the cellular model by
assuming an appropriate {\sl ansatz} for the probability density -- usually a
sum of Gaussians -- which then leads to a self-consistency relation determined
by the requirement that probability densities of all particles be the same,
thereby fixing the parameters of the {\sl ansatz}.~\cite{Hone83} Here we follow
the spirit of this approach but we note that in the case of a system interacting
with flat potential such as HCSS, the Gaussian ansatz may not be adequate.
Moreover, in the A15 lattice the probability distribution of particles in
columnar positions should be very anisotropic and in fact nonspherical. Given
that our main task here is to examine the possibility that the phase diagram of
the HCSS system includes the A15 lattice, it seems appropriate to capture the
geometrical details of the different structures by calculating the configuration
integral numerically.

Our procedure places the central particle within a cage of its neighbors that
are themselves allowed to sample a distribution of positions such that the
probability densities of all particles at equivalent sites are the same. This is
achieved by sampling the actual probability density of the central particle
continuously, and recreating identical distributions at the equivalent neighbor
positions.~\cite{MCremark} After equilibrating the system for a certain number
of steps and checking for the consistency of the distributions of the central 
particle and the neighbors, we calculate the configuration integral and the free 
energy. This approach could be dubbed a ``single-particle Monte Carlo method'';
we note, however, that it is in fact a numerical variant of the self-consistent
mean-field approximation and a relative of the variational method proposed by
Kirkwood~\cite{Kirkwood50} rather than a simplified variant of a true
multiparticle Monte Carlo integration. Our procedure is single particle because
we approximate the multiparticle probability density by a product of 
single-particle probability densities and thus ignore long-wavelength
excitations.

To ensure that the model reduces to the quantitatively successful cellular 
free-volume theory used in Section II~\cite{Curtin87}, we modify the HCSS
interparticle potential by not allowing the hard core of the particle to leave
the cell. This gives the correct free energy for small shoulders, 
($\epsilon<k_{\scriptscriptstyle B}T$) in good agreement with Monte Carlo
analyses,~\cite{Rascon97,Velasco98} as well as for large shoulders, where the
particles behave essentially as hard spheres moving in a constant potential. We
expect that between these two limiting cases, the correlations between nearest
neighbors (and thus the free energy) should also be well approximated. While 
this method can not include the fluid phase, we are really interested in the 
stability of the A15 lattice so that the only relevant aspect of the fluid-solid 
transition is melting. Instead of going into the details of the fluid HCSS phase 
and analyzing the transition with, for example, 
Ramakrishnan-Yussouff theory,~\cite{Ramakrishnan79} we can estimate the melting
curve by extrapolating the zero-temperature FCC melting point, which is, as we
will see, far enough from the other phase boundaries to suggest that the latter
persist within the temperature range that we study.

\begin{figure}

\centerline{\epsfig{file=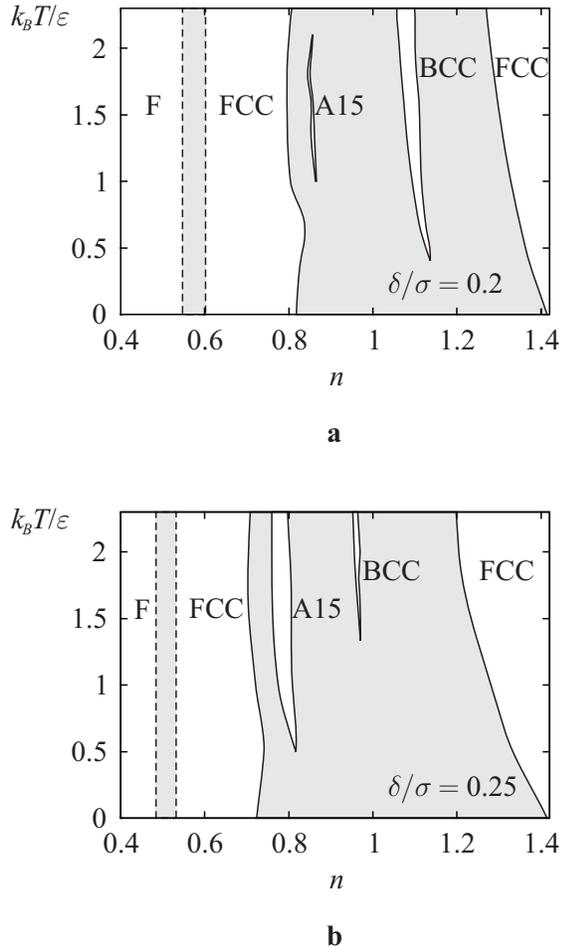}}

\vspace*{6mm}

\caption{Phase diagram of the hard-sphere--soft-shoulder potential for various
shoulder widths: (a) $\delta=0.2$ and (b) $\delta=0.25$. For these widths there
is no expanded BCC lattice. The A15 lattice appears as one of the intermediate
structures, interposed between the expanded and condensed FCC structures. The
A15 islands of stability are elongated along the temperature axis, indicating
that this structure should be sought after by varying the density rather than
temperature. Note that the single-phase regions are small compared to
coexistence regions. The dashed lines indicate the zero-temperature expanded
FCC--fluid (F) coexistence.}
\label{phasediagrama}
\end{figure}
\begin{figure}

\centerline{\epsfig{file=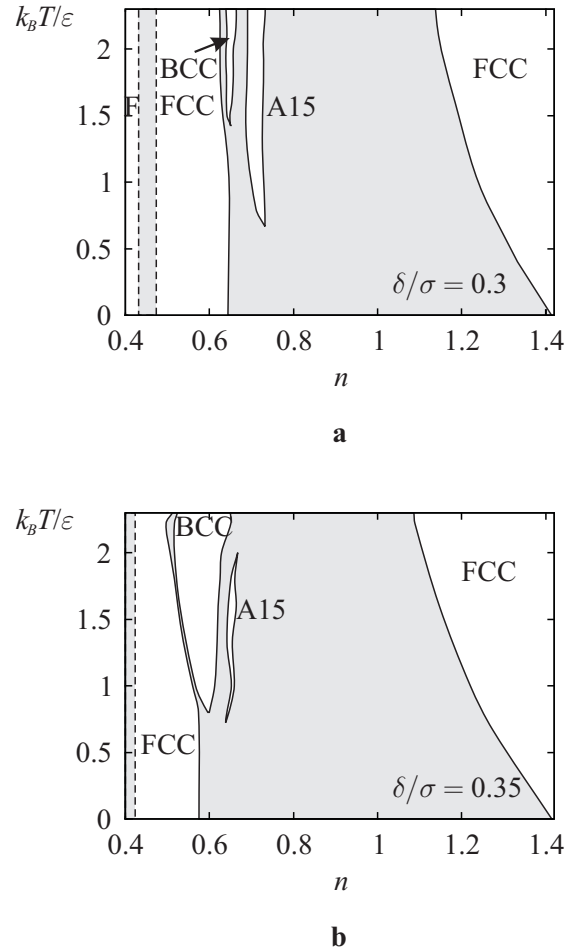}}

\vspace*{6mm}

\caption{Phase diagram of the hard-sphere--soft-shoulder potential for various
shoulder widths: (a) $\delta=0.3$ and (b) $\delta=0.35$. For these widths there
is no condensed BCC lattice. As in the previous figure, the A15 lattice appears 
as one of the intermediate structures, interposed between the expanded and 
condensed FCC structures.}
\label{phasediagramb}
\end{figure}

Using this approach, we have analyzed the solid part of the phase diagram at low
temperatures, focusing on the FCC, BCC, and A15 lattices. The coexistence
between the phases was determined using the Maxwell double-tangent construction,
and the results are shown in Figs.~\ref{phasediagrama} and~\ref{phasediagramb} 
for shoulder widths $\delta/\sigma=0.2,0.25,0.3,$ and 
0.35.

The main features of the phase diagram can be summarized as follows:

(i) We find that the A15 lattice can be a stable state of the HCSS system
between the expanded and condensed FCC phase. As expected from the structure of
the pair
distribution function, this lattice is stable for shoulders neither too
narrow (which would make the particles too similar to ordinary hard spheres) nor
too broad (which would destroy the comparative advantage of the A15 lattice over
the FCC and BCC lattices). The minimal and the maximal shoulder widths roughly
correspond to $\delta=0.2$ and 0.35, respectively. Note that these widths are
consistent with the structural parameters of the dendrimer compound where this
lattice was observed experimentally, although the effective hard core of the
micelles may not necessarily coincide with the benzyl inner segment of the
dendrimers.

Temperature is also a crucial parameter of the stability of the A15 lattice and
should not be too low nor too high. The A15 island in the phase diagram appears
to be centered around $k_{\scriptscriptstyle B}T/\epsilon\approx1.4$ and its
temperature range spans about $2k_{\scriptscriptstyle B}T/\epsilon$ at most, the
maximum being at $\delta\approx0.27$. The A15 lattice is a delicate structure:
for all $\delta$ and $T$, the pure A15 structure occurs only within a rather
narrow range of density centered at $n\approx0.85$ with width
$\Delta n\approx0.05$.

(ii) In the density-temperature plane, islands of stability of the 
intermediate-density solid phases ({\sl i.e.}, expanded FCC, expanded BCC, A15,
and condensed BCC lattice) are elongated along the temperature axis, implying
that over broad ranges of temperature, the phase sequence does not change
dramatically with density. This departs from the low-temperature extrapolations
of the results of perturbative theory:~\cite{Velasco98} the stripe-like topology
of the phase diagrams presented in Fig.~\ref{phasediagrama} and 
Fig.~\ref{phasediagramb} is free of critical and/or triple points as well as the 
corresponding isostructural transitions. This is most likely a consequence of 
the rather broad shoulder.

(iii) For shoulders as broad as necessary to stabilize the A15 lattice, the
phase diagram is characterized by relatively small islands of stability of the
intermediate phases: two-phase coexistence dominates the phase diagram. This
feature appears to be specific to convex interparticle potentials, whereas in
case of concave potentials the regions of coexistence are typically much
narrower.~\cite{Lang00}

While the stability of the A15 lattice is certainly not limited to HCSS
potentials, our findings indicate that at least in some systems it could coexist
with FCC, BCC, or perhaps another lattice over a rather wide density range. As
far as the structural identification of the samples is concerned, this simple
fact may have important consequences. In the case of phase coexistence, the
interpretation of X-ray measurements is difficult. In practice, the space groups
of the coexisting phases can only be determined unambiguously by varying the
external parameters and moving from one island of stability across the
coexistence region to the other island of stability. Given the stripe-like
topology of the phase diagram, the parameter to be varied should be the density
and not temperature. Even with this proviso, the identification can be difficult
because of a considerable overlap between the diffraction peaks of the
structures involved. For example, the difference between the patterns of A15 and
BCC lattice is not very striking -- the reflections being at
$\sqrt{2},\sqrt{4},\sqrt{5},\sqrt{6},\ldots$ and at
$\sqrt{2},\sqrt{4},\sqrt{6},\ldots$ for the A15 and BCC lattices, 
respectively~\cite{Hahn83} -- and could be masked by the form factor of the
particles, which usually falls off quite rapidly with the 
wavevector.~\cite{McConnell93} These shortcomings could be overcome by
complementing the X-ray studies by calorimetric measurements. 

Last but not least, we note that the relative rarity of the A15 lattice in real
colloidal crystals may be caused by other, non-equilibrium mechanisms that can
slow down the formation of a pure A15 colloidal crystal. For example, it is
conceivable that the relatively simpler BCC lattice is kinetically favored over
the A15 lattice in experiments which evaporate solvent to form crystallites. 
Given that the free energies of the different lattices are typically rather 
small, the equilibration can take very long and the actual ground state may be
hard to observe.

Nevertheless, along with recent experimental~\cite{Balagurusamy97} as well as 
theoretical studies~\cite{Watzlawek99} our preliminary results point to the 
necessity of extending the phase diagram of soft spherical particles by the A15
lattice and possibly other non-close-packed lattices. At the same time, the 
more intricate phase diagram may be more difficult to determine and understand.

\section{Conclusions}

In this study, we have extended the geometrical interpretation of the free
energy of weakly interacting classical particles, and we have complemented the
well-known close-packing rule with a minimal-area rule. The incompatibility of
the two principles leads to frustration which gives rise to range of possible
equilibrium structures, depending on the relative weight of the two terms. Our
proposed scheme is a computationally simple, zeroth-order description of 
soft-sphere crystals which is more transparent than detailed numerical models,
most notably molecular modeling.

As such, our theory provides a robust insight into the self-organization of such
objects, which should be useful for the engineering of colloidal crystals. The
relevance of these universal guidelines is as broad as the use of colloids
themselves, ranging from photonic bandgap crystals~\cite{Tarhan96,Busch98} to
micro- or mesoporous materials used for chemical microreactors and molecular
sieves.~\cite{Jenekhe99,Sakamoto00} To meet the demands of a particular
application, these designer materials must be characterized by a given lattice
constant, symmetry, and mechanical properties, and, in the case of porous
structures, void size and connectivity. All these parameters can be controlled
by tuning the structure and size of the (template) colloidal particles and the
interaction between them, and our model establishes a semi-quantitative
relationship between particle geometry and bulk material properties.

We envision this work to be extended in several directions. One problem to be
addressed is to locate the A15 lattice within the phase diagram using Monte
Carlo analysis, starting with a square-shoulder potential but also employing
less generic short-range potentials. In addition, we will further explore the
analogy between colloidal crystals and soap froths in view of the geometrical
approach that we have adopted here. Another interesting aspect of future
work could be to use the model to derive some of the mechanical properties of
colloidal crystals, such as the shear and Young moduli. One could also study the
stability of non-cubic lattices, which have been mostly neglected so far, in an
attempt to understand non-spherical colloidal particles. Work along these lines
should lead to easily verifiable predictions and a deeper insight into the
physics of colloids.

\acknowledgments

We gratefully acknowledge stimulating conversations with M.~Clerc-Imp\'eror,
G.H.~Fredrickson, W.M.~Gelbart, P.A.~Heiney, C.N.~Likos, T.C.~Lubensky, 
V.~Percec, J.-F.~Sadoc, and A.G.~Yodh.

This work was supported in part by NSF Grants DMR97-32963, DMR00-79909 and
INT99-10017, the Donors of the Petroleum Research Fund, administered by the
American Chemical Society, and a gift from L.J.~Bernstein. R.D.K. was also
supported by the Alfred~P.~Sloan Foundation.


\begin{references}
\bibitem{MCremark} Technically, this was done by storing the accepted moves of
the central particle and then assigning randomly selected accepted moves to its
neighbors. The array of accepted moves was big enough so that the free energy 
did not depend on its size, and was updated continuously. After equilibration,
this leaded to identical probability distributions at all equivalent positions.
\end{references}
\end{document}